# BRINGING SIMULATION TO APPLICATION : PRESENTATION OF A GLOBAL APPROACH IN THE DESIGN OF PASSIVE SOLAR BUILDINGS UNDER HUMID TROPICAL CLIMATES

F. GARDE,* H. BOYER,* R. CELAIRE**

Laboratoire de Génie Industriel, University of Reunion Island, IUT de Saint Pierre,
40 avenue de Soweto, Saint-Pierre, Reunion Island, 97 410, France,
(+262) 96 28 91, (+262) 96 28 99, garde@univ-reunion.fr
** Concept Energie, 1 rue Mirabeau, Lambesc, 13410, France,
(+33) 04 42 92 84 19, (+33) 04 42 92 71 36, Robert.Celaire@wanadoo.fr

**Abstract** – In early 1995, a DSM pilot initiative has been launched in the French islands of Guadeloupe and Reunion through a partnership between several public and private partners (the French Public Utility EDF, the University of Reunion Island, low cost housing companies, architects, energy consultants, etc...) to set up standards to improve thermal design of new residential buildings in tropical climates. This partnership led to defining optimized bio-climatic urban planning and architectural designs featuring the use of passive cooling architectural principles (solar shading, natural ventilation) and components, as well as energy efficient systems and technologies. The design and sizing of each architectural component on internal thermal comfort in building has been assessed with a validated thermal and airflow building simulation software (CODYRUN). These technical specifications have been edited in a reference document which has been used to build over 300 new pilot dwellings through the years 1996-1998 in Reunion Island and in Guadeloupe. An experimental monitoring has been made in these first ECODOM dwellings in 1998 and 1999. It will result in experimental validation of impact of the passive cooling strategies on thermal comfort of occupants leading to modify specifications if necessary. The paper present all the methodology used for the elaboration of ECODOM, from the simulations to the experimental results. This follow up is important, as the setting up of the ECODOM standard will be the first step towards the setting up of thermal regulations in the French overseas territories, by the year 2002.

## 1. INTRODUCTION

Each year 20,000 dwellings are built in the French Overseas Territories. There are four French Overseas Territories (called DOM) : two islands are located in the Caribbean (Martinique and Guadeloupe), one island is situated 400 km to the east of Madagascar in the Indian Ocean (Reunion Island) while the fourth DOM (French Guyana) is in South America North of Brazil. Each one of these DOM experiences a hot and humid climate, tropical for the islands of Guadeloupe, Martinique and Reunion and equatorial for French Guyana. Two thirds of these new dwellings are social housing. Initially these new dwellings were built without air conditioning or hot water. This situation has led to haphazard installation of instant electrical hot water boilers and poorly located, designed, sized, and maintained individual air conditioning systems. The lack of building codes on thermal design of buildings, combined with designs widely inspired from temperate climates together with economic constraints of tight budgets for public constructions have led to the development of buildings totally unfitted to tropical climate. Large population increase in these DOM, rise in living standards, and the decreasing costs of air conditioning systems result in a real energy, economical and environmental problem.

When considering economical aspects, high electricity production cost also creates continual high deficit for the French Electricity Utility (EDF). EDF loses 350 millions Euros every year in overseas territories. The average selling price of electricity is less than 9 Euro cents (less then half the real production-distribution cost) because of the French pricing policy (selling price of electricity is the same as in mainland France).

All these factors point out that passive solar cooling for thermal design of buildings is of great economical, social and environmental relevance. A long term overall program to improve comfort and energy performance in residential and commercial buildings has been set up in these DOM. In the new housing sector, a quality standard label has been launched concerning the building envelope, hot water production systems and air conditioning systems and components.

## 2. THE ECODOM STANDARD

A Demand Side Management pilot initiative called ECODOM was launched in early 1995 in the French islands of Guadeloupe and Reunion through a partnership between the French electricity utility (EDF), the French authority for environment and energy conservation (ADEME), the ministries of Housing, of Industry and of Overseas Territories, the University of Reunion island and several other public and private partners, such as low cost housing companies, architects, energy consultants, etc...

### 2.1 The objectives

ECODOM standard aims to facilitate construction of naturally ventilated comfortable dwellings whilst avoiding the usual necessity of electricity driven compression air cooling. The ECODOM standard provides simple technical specifications to do so, at an affordable price.

Existing bibliography on passive thermal design of buildings is extremely rich and varied. Publications often focus either on the optimization of one component of building (Bansal, 1992), (Malama, 1996), (Rousseau 1996) (Peuportier, 1995) or on the presentation of a performing bio-climatic project (Filippin et al, 1998), (Ashley and Reynolds, 1993) or on global approach of building with a description of the passive solar strategies to implement in it (Garg, 1991), (Hassid, 1985), (Millet, 1988), (Gandemer 1992). These publications often have an obvious interest for building physicists but cannot be easily applicable in a national overall program to generalize improvement of thermal performances of buildings : because scientific preoccupations are often far from economical building reality. In addition architects and engineers lack of time to learn to use scientific tools and to read research reports in the fields of thermal design of building. Thus, to bridge the gap between building designers and building physicists, a simple, straightforward and pedagogical language must be spoken.

Another objective of the ECODOM project was to define simple rules which can be easily understood by the whole building community.

Finally, ECODOM has also social objectives because more than half of the 20 000 dwellings in the French overseas territories are built by public housing companies. ECODOM allows to give basic thermal comfort to people who will never be able to afford air conditioning investments and subsequent electricity bills.

A reference document comprising a pedagogical presentation of the technical passive options was published in 1996. This document enables all building designers to speak the same language in the study of future ECODOM buildings.

Once the document was made, the next step was obviously to implement the standard to 300 pilot new dwellings throughout the years 1996-1998, then, after a technical and sociological validation period, to expand this pilot phase to the residential sector on a much broader scale (objective of 2000 new dwellings per year), and complete similar global energy efficiency projects to existing housing and to large and medium size commercial buildings. Then, the experimental feed back will allow the setting up of thermal regulations in the new housing sector by the year 2002 (see figure 1).

*Fig. 1 : The ECODOM process*

**2.2 <u>The prescriptions</u>**

Comfort level is reached through an architectural building design adapted to local climate : the dwelling is protected from the negative climatic parameters (mainly the sun) and benefits from positive climatic factors (the wind).

Achievement of a good level of thermal comfort requires respecting various compulsary rules. These rules concern immediate surroundings of the dwelling and envelope components. These rules cover five issues:

1) Location on site (vegetation around the building);

2) Solar protection (roof, walls, windows);

3) Natural ventilation (exploitation of trade winds, and optimized ratio of inside/outside air-permeability of dwelling envelope) or mechanical ventilation (ceiling fans);

4) Domestic hot water production ( solar water heaters , efficient gas water heaters or properly designed off-peak hours servo-controlled high efficiency storage electric water heaters);

5) As an option, high efficiency intermittent air conditioning systems for bedrooms (air tightness of rooms, efficient components, controls,..).

## 3. METHODOLOGY

To define these quality standards specifications, a large number of simulations were computed for each component of building envelope in order to quantify thermal and energetic impact of each one on thermal comfort within the building. Various authors have already worked on specific issues concerning the outside structure of the building : (Bansal, 92) on the effect of external color, (Malama, 96) on passive cooling strategies for roofs and walls, (Rousseau, 96) on the effect of natural ventilation, (De Walls 93) on global considerations on the building adapted for a defined climate.

Our approach consisted in studying typical dwellings, selected as the most representative types of the ones built in the Reunion island, in terms of architecture and building materials (see figure 2).

*Fig. 2 : Typical dwelling*

Simulations were carried out with the use of a building thermal and airflow simulation software on envelope components (roof, walls, windows) and on natural ventilation, in a way allowing to estimate the influence of each one of the above specifications in terms of thermal comfort and energetic performances. This has led to the definition of efficient passive technical specifications for each comprising part of building envelope and likewise to a minimum porosity ratio for optimizing natural ventilation. These simulations, their analysis and synthetic results have been presented in (Garde, 99a) and (Garde 99b).

## 4. ECODOM : THE REFERENCE DOCUMENT

We will illustrate here after some of the final results which are presented in the ECODOM reference document (Celaire, 1997). This reference document is essential since it is used as a common tool for study and design of new ECODOM projects by architects, building physicists and engineers.

### 4.1 Solar protection

In humid tropical climate, source of uncomfort result from internal temperature increase due to poor architectural design, primarily due to lack of solar protection (shading). 80% of temperature rises are due to solar radiation and the remaining to conduction exchanges. Setting up efficient solar protection specifications is the second fundamental step in thermal design of buildings. This protection concerns all external components of dwelling envelope : roof, walls and windows.

#### Solar protection of the roof

Roofs can account for up to 60% of overall envelope heat gains in dwellings. Efficient solar protection of roofs is therefore of prime necessity for optimal thermal design.

The following table is valid for terraced roofs, tilted roofs without attics, or roofs with closed or barely ventilated attics.

Table 1 only gives results for polystyrene and polyurethane, as these types of insulation are the most commonly used at an affordable price in humid tropical climates. Other alternatives exist (low emissivity materials for instance) and can be used if their equivalent thermal resistance is sufficient compared to values given in Table 1.

*Table 1 : Roof solar protection*

**Solar protection of walls**

Thermal gains from walls account for 20 to 30% (40 à 65 % for dwellings which are not under roof) of overall envelope heat gains in dwellings. Various solutions enable efficient solar protection of walls from sunlight : horizontal or vertical overhangs or sun-breakers, thermal insulation. Results obtained from simulations lead to the table 2 which give, for Reunion island, optimum dimensions of overhang according to wall orientation and to wall thermal resistance.

*Table 2 : Overhang - minimum d/h ratio values to be respected.*

*Table 3 : Insulation of walls (in cm) for different orientations and external colours (for a conductivity of 0.041 W/m.K)*

When walls do not have overhangs or shading systems, minimum insulation thickness (in cm) required for the various types of walls and various orientations are shown in table 3.

If values of d/h seem excessive, other alternatives such as vertical shading systems or double-skin systems with ventilated air-gaps may be considered. In that specific case, no sizing specifications are needed for each wall orientation (shading effect is sufficient whatever the orientation is provided the shading component covers the entire wall surface). Meanwhile, these solar shading principles and sizing values are fully compatible with the creole architectural components such as verandas and balconies which insure efficient solar protection of dwellings walls and windows together with creating outdoor living spaces.

**Solar protection of windows :**

Solar protection of windows is essential, not only because they account for 15 to 30% of overall envelope heat gains in dwellings but also because they contribute to increasing uncomfort experienced by occupant, due to the instant heating of ambient air temperature and an exposure to direct or reflected sunlight. All windows must therefore be protected by some kind of window shading system, such as horizontal overhangs and other shading devices such as venitian blinds or opaque, mobile louvers (see Fig. 3). Simulations enabled to optimize geometric characteristics of horizontal overhangs in relationship with glazing. (see table 4).

*Table 4 : Values of d/(2a+h) (case 1), or d/h (case 2)*

*Fig. 3 : Some of the shading devices required by ECODOM*

**4.2 Natural ventilation**

In warm climates, natural ventilation is the most usual means of cooling both occupants and buildings :

Natural ventilation, depending on flow-rate values (or air change per hour), can ensure three functions :

- Weak flow-rate (1 to 2 ACH) : preservation of hygiene conditions in building by internal air renewal;
- Moderate flow-rate (40 ACH): dissipation of internal heat gains and cooling of building envelope;
- High flow-rate (more than 100 ACH) : improvement of comfort of occupants by increasing heat transfer at skin level .

High air speed and even distribution of air flow through dwelling increases sudation process. This is the only means enabling to create cooling by compensating for simultaneous high temperatures and high hygrometry.

Our aim is therefore to find exterior/interior walls porosity coupling enabling to reach 40 ACH. On one hand dwelling envelope will be sufficiently cooled and on the other hand, such ACH values allows to benefit from wind speeds of 0.2 à 0.5m.s$^{-1}$, which is largely sufficient, when taking into account climatic parameters (outside temperature seldom greater than 32°C), to insure good comfort level.

Simulations run in (Garde, 99) and (Garde, 97) showed that critical rate of 40 ACH is obtained for designs with minimal exterior and interior porosities of 25% whether building is light or high mass. Natural ventilation is simply more effective during night time for heavy structures, whereas in light structures it serves mainly to evacuate daytime overheating.

Thus dwellings should have layouts allowing for effective cross ventilation of all rooms (see Fig. 4). At each level or floor, there should exist openings in each main room, on at least two opposite facades (main rooms being bedrooms and living, dining room and other living spaces). Interior lay-out should also be designed to allow outside air to flow through main rooms from one facade to the other through halls, corridors, doors and other internal openings in room partitions.

*Fig. 4 : Cross ventilated dwelling*

Calculation details needed to determine that exterior and interior porosities are greater than 25% are reported below. First thing to do is to calculate the mean porosity of two opposite facades of dwelling. Internal and external porosity required by ECODOM standards is 25% of the mean.

$$P1 = \frac{So1}{Sp} \geq 0.25$$

So1 : Net surface area of external openings, main rooms (façade 1);

$$P2 = \frac{So}{Sp} \geq 0.25$$

So2 : Net surface area of external openings, main rooms (façade 2);

$$Sp = \frac{Sp1 + Sp2}{2}$$

Sp1, Sp2 : Total surface area of façades 1 and2 of main rooms;

$$Si1 \geq So1 \text{ or } So2$$
$$Si2 \geq So1 \text{ or } So2$$

Si1 and Si2 : Total surface area of internal openings.

**4.3 Domestic hot water**

It is essential that the dwellings are equipped with energy efficient long-lasting and economic, domestic hot water heating systems. Water heating can be solar, electric or gas or a combination of such systems.

In the case of solar water heaters for example, the system must conform technical control by the french centre for scientific and technical building studies (C.S.T.B.). Total minimum solar collectors area should be defined in relationship with the size of dwelling.(see table 5). Capacity of water storage tank should be 60 to 120 liters per square meter of net collector area. Conventional minimum annual productivity should be 700kWh per net square meter of the collector area.

*Table 5 : Technical characteristics - solar water heaters*

As far as electrical heaters are concerned, they must feature an approved French standard manufacturing seal (Norme Française). Minimal capacity of water heater and maximum value of cooling constant (defining storage efficiency of heaters), depending on number of main rooms of dwelling, are required.

## 5. PRESENTATION OF AN ECODOM PROJECT : "LA DECOUVERTE"

At present time, three experimental operations have been built in Reunion Island, and two in Guadeloupe, what represents a total number of 300 dwellings.

One of these called "La Découverte" is represented by Fig. 5.

44 dwellings were studied with the ECODOM specifications as part of "La Découverte" project.

*Fig. 5 : ECODOM "La Découverte" - the original project.*

This project started in 1996 when first contacts were taken with project architect and the building owner.

The main problem was to create confident working relationship between physicits and engineers in charge of promoting Ecodom concept and project architect : to be successful ECODOM design approach had to be understood not as factor limiting architectural creativity.

Figure 6 illustrates project after application of ECODOM specifications.

*Fig. 6 : Modifications of the original project*

The ECODOM modifications applied to project are explained here after.

No modification was made for site location because gardens around buildings had already been planned. In addition project has an optimal orientation as main façades have a north and south orientation, with limited west and east walls exposure to morning and afternoon low incidence sunrays. Thus the only façade to be upgraded was the northern one as the southern one is seldom exposed (for the southern hemisphere latitude of Reunion Island, midday sun shines on southern façade only in december).

As far as solar protection of roof was concerned, architect had already planned 5 cm insulation for red and blue roofs. According to table 1, it is not enough. Therefore, he was asked to increase insulation thickness up to 8 cm (value for medium colors) and to use polystyrene as insulation rather than mineral wool. Mineral wool is fairly cheap but not very well adapted to tropical climates : it loses its thermal properties when it absorbs ambient humidity .

As far as solar protection of walls is concerned, the color is light and materials used are hollow concrete blocks. In table 2, it is specified to install 1 cm insulation on East and West sides. The southern façade was not insulated because it is protected by overhangs. In addition as the initial project didn't provide solar protection for bedrooms windows, is was specified to add overhangs on exposed windows of northern facade.

However, the solar protection of windows is strongly correlated to the natural ventilation. The bigger the opening, the wider the overhang must be.

Concerning natural ventilation, dwelling porosity in initial project was not sufficient. For example, ECODOM standards required for bedrooms a 2 square meters porosity while the initial project only had 1.44 square meter. This is why all initial windows have been replaced by glazed doors in order to increase porosity. To reach the same porosity indoor,it was asked to add fanlights above the bedroom doors . Balconies were also added all along the facade for solar protection of windows and of walls (see figure 7).

*Fig. 7 : Floor plan of an ECODOM dwelling. Increase in porosity of bedrooms (glazed windows) and solar protection of northern facade (balconies).*

All the modifications expressed above have been taken under considerations by the architect and integrated into the final project which is represented by Fig.8. The operation « La Découverte » ended in April, 1999, which is three years after our first contacts with the persons responsible for the project. We also got through a phase of technical support in the thermal conception to a phase of implementation of the project. To make the methodoly completed, it only lacks the experimental monitoring and feed back, what is going to be presented in the following section.

*Fig. 8 : The final project (april 1999)*

## 6. THE TECHNICAL AND SOCIOLOGICAL MONITORING

### 6.1 Description

An experimental monitoring has been launched for the first ECODOM dwellings in order to validate experimentally impact of the passive cooling specifications on the performances of the envelope of the building and on the comfort of occupants.

The campaigns of instrumentation occured on the first two operations ECODOM implemented in Reunion : the first one has been presented in this paper and the second one was built one year earlier and is called the operation «La Trinité».

Our objective was firstly to test the performances of the buildings without the occupants. So, we could make our own scenarios of measure and modify the configurations of apartments without disturbing the tenants. Secondly, we led a campaign of measure in occupied flats (for the Trinity only) to test the dwellings in real case.

The measurements results and analysis have been supplemented for the both projects with a sociological study in inhabitated dwellings in order to understand how thermal comfort, acoustic comfort (in tropical climates, it might be difficult to obtain thermal comfort and acoustic comfort at the same time), visual comfort and dwelling environment are appreciated by occupants

Table 6 gives the schedule of the campaigns of measures for the two operations.

*Table 6 : Experimental schedule for non occupied and occupied periods*

These campaigns happened during the two last warm seasons (on 1998 for « The Trinity » and on 1999 for « The Discovery ») for estimating the thermal performances of flats during extreme climatic periods (the warm and wet season in Reunion occurs from the middle of December to the middle of April). The sociological study was led for the operation « The Trinity » in April, 1998. It is also foreseen for « La Découvert » by April, 2000.

With the agreement of the housing companies, in each case ("La Trinité" and "La Découverte") four dwellings have been instrumented during the hot season. Dwellings have been selected in order to allow comparison and studies of various parameters such as roof and walls thermal insulation, natural ventilation. Thus we chose a flat under roof (et en pignon) and the same flat in a lower level to test the influence of the roof and of the solar protection of the glazings . The two other instrumented flats are rigorously identical in orientation and in level or floor, this to test different scenarios such as the effect of natural ventilation with a flat which serves us as a reference and the second on whom we modify its configuration.

A portable data-logger was set up close to the buildings to record all the climatic data (external temperature and humidity, direct and diffuse solar radiation, wind speed and wind direction).

During non-occupied period, sensors used are thermocouples for measurement of indoor air temperature, resultant temperature and surface temperatures on walls and roofs; thermo-hygrometers for the relative humidity, accurate anemometer for indoor air speed (see fig.9). Data are collected every minute and the average is made every hour.

During occupied period, sensors are like small "white boxes" with internal memory which can be wall mounted. They give temperature and hygrometry measurements each 30 minutes during several months.

*Fig. 9 : Measurements of indoor air and resultant temperature inside dwelling*

## 6.2 <u>Presentation of some experimental results</u>

We shall present in this section only some significant experimental results. The whole results and their analysis are contained in two research reports (one for the operation "La Trinité" and one for the operation « La Découverte ») available with the authors.

*Impact of the solar protection of roof :*

The insulation of the roof is a first urgency in solar protection within the framework of the ECODOM process. To put in evidence the influence of the roof, we have instrumented two identical apartments (same as the one figure 8) but with one under roof and the another one at a lower level. In that case, when the two apartments are closed, only the energetic contribution of the roof is responsible of the gap between the indoor resultant temperature.

Figures 10 and 11 show respectively the insolation during the period of the instrumentation and the evolution of the indoor resultant temperature of the two apartments. A systematic offset is observed of 1¨C to 1,5¨C between the two apartments during the whole period of measure. One finds again the same type of result for the two campaigns of measure ("La Trinité" and "La Découverte") : a flat under roof with a good insulation corresponding to ECODOM has always his resultant temperature upper of 1¨C to 1,5¨C compared to the flats which are not under roof. With a badly designed roof, distances of more of 3¨C have been observed. This type of measure confirms the major importance of a good insulation of the roof in tropical climate.

*Fig. 10 : Solar radiation during the assessment of the solar protection of the roof*

*Fig. 11 : Comparison of the resultant temperature in two dwellings (one under roof and the other one at intermediate level*

*Occupied period (operation "La Trinité")*

We present here the results obtained for a occupied flat of the operation « The Trinity ». The period of instrumentation lasted two months (in March and April, 1998). The instrumented flat ihad two levels. The first level had as two bedrooms, the kitchen and the dining room. The upper level under roof had only a third bedroom. This flat had two infringements compared to the ECODOM specifications : the first one concerned the insulation of roof which was only 5 cms while ECODOM asked for 10 cms (the roof was of grey colour). The second one concerned the ventilation of the bedroom under roof : the percentage of ventilation was sharply lower than the percentage required by ECODOM.

Figures 12 and 13 show the set of the temperature/humidity couples over the period of instrumentation for a room (chamber 2) of the first level and for the room under roof (chamber 3).

We observed generally that the rooms which are not under roof and well naturally ventilated have globally the same thermal and airflow behavior which is represented by Fig. 12. We are on average 10 % of time outside the zone of comfort

The room (bedroom 3) under roof is very warm. The distribution of the points on the psychrometric chart is very spread over the axis of the resulatant temperatures. Although this room is naturally ventilated and is often used by the tenant, we are 40% of the time outside the comfort zone. It is not so the way of living that influences the thermal behavior of the room, but rather a problem of the insufficient heat insulation.

Besides, we can suppose that the crossing ventilation of the room is apparently insufficient.

*Fig. 12 : Temperature/humidity couples of an occupied flat, room at intermediate level*

*Fig. 13 : Temperature/humidity couples of an occupied flat, room under roof*

**6.3 Synthesis**

Experimental and sociological results will enable to have feed back informations about ECODOM dwellings and to supply eventual appropriated corrections to quantitative values of passive solar specifications. Above all, experimental feed-back allow us to know if dwellings are well adapted to occupant way of life.

We can't point out the whole points of these studies but some of the main conclusions can be indicated.

- The solar protection of the roof stays one of the main point in the thermal design of buildings in tropical climate. We were able to show some defaults of the implemented insulation in the experimented operations leading important overheating. It is also necessary to maintain the specifications of table 1 concerning the minimum thickness of insulation needed.

- Natural ventilation showed itself effective in the tested flats except at night in the case of very weak night breezes where an overheating of 1¨C was observed compared to the outside temperature. Also, the ECODOM percentage of openings of 25 % is a minimum value below which we can not get through.
- Finally, the sociological study showed that the tenants were globally satisfied with their flat considering the thermal aspects. However the dissatisfaction resulted from the bad acoustic design of the flats. And we touch here one of the major problems of the conception in tropical climate : the duality of thermal and acoustic design. One the one hand, we must open the dwelling to look for thermal comfort, on the other hand, the acoustic comfort is reached by closing the windows.

## 7. CONCLUSION

Methodology used for developing ECODOM standards, from simulations to experimental results, has been presented.

This whole work led to a pedagogical reference document defining efficient passive solar cooling strategies and specifications for each component of dwelling envelope and likewise minimal porosity ratios to optimize natural ventilation. Dwellings to be constructed according to ECODOM standards should follow these specifications. ECODOM actors (architects, designers, engineers, building physicists,…) now speak the same "language" by using the reference ECODOM document.

An experimental monitoring has been set up for first ECODOM dwellings constructed in order to validate experimentally impact of passive cooling specifications on comfort of occupants. This experimental work has allowed a real scale feed back. Such monitoring is essential, as the ECODOM standard is the first step towards the setting up of building codes and regulations on energy efficiency in envelope of residential buildings in the French overseas territories, that should be effective by the year 2002.

| | |
|---|---|
| 1994 | Identification of the energetic, climatic and social contexts |
| 1995 | Setting up of a network between the ECODOM partners (ministry of housing, architects, the french autority for environment and energy conservation, energy consultants, the university of Reunion island, low cost housing companies, public utilities…)<br>Choice of typical dwellings and of a typical day for running the simulations<br>Simulations on typical dwellings with the thermal and airflow software CODYRUN |
| 1996 | Definition of the performant passive cooling solutions.<br>Edition of the ECODOM reference document after numerous in and outs among the different partners for taking into account the pedagogical aspects of the document.<br>Search for the first ECODOM projects. Communication. |
| 1997-99 | 300 ECODOM dwellings built in the DOM. 5 experimental ECODOM operations. (2 in Guadeloupe and 3 in La Réunion)<br>Experimental and social follow up.<br>Consulting and research of new projects. |
| 2000 | Experimental feed back and modification of the reference document. |
| 2002 | Setting up of building codes and regulations on energy efficiency in envelope of residential buildings in the French overseas territories. |

Fig. 1 : The ECODOM process

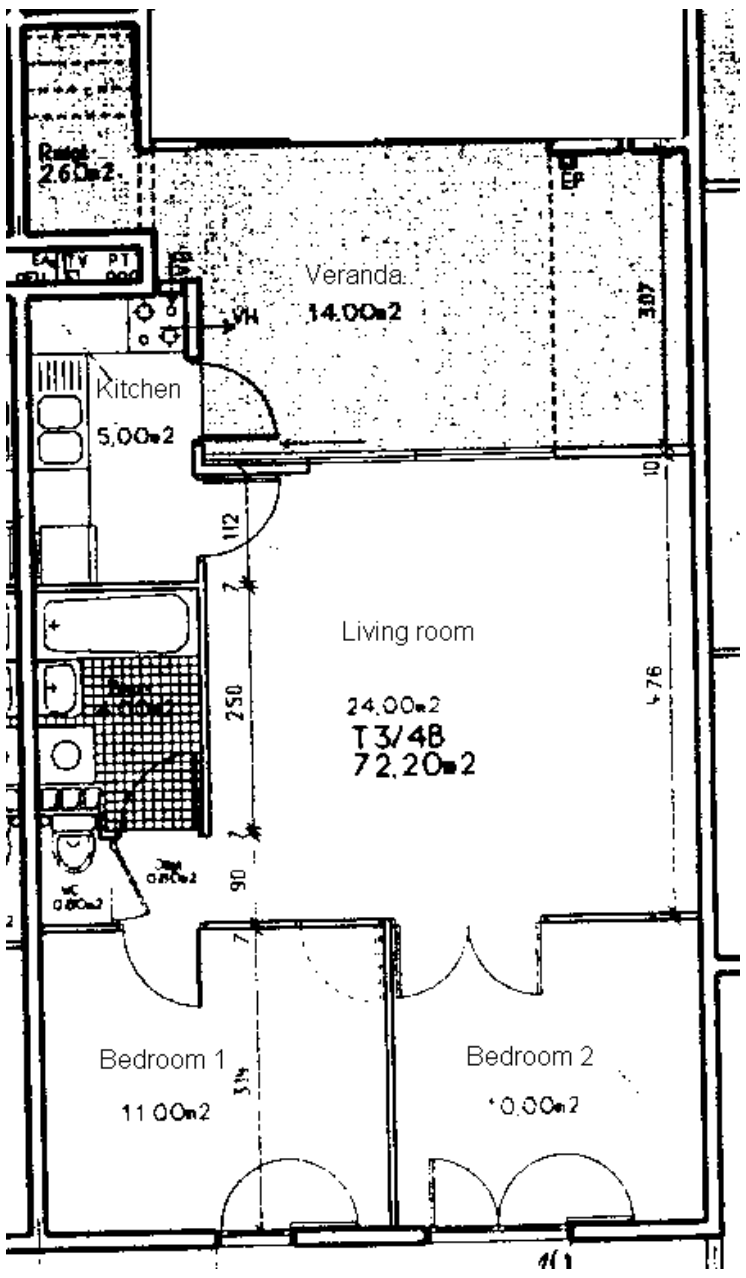

Fig. 2 : Typical dwelling

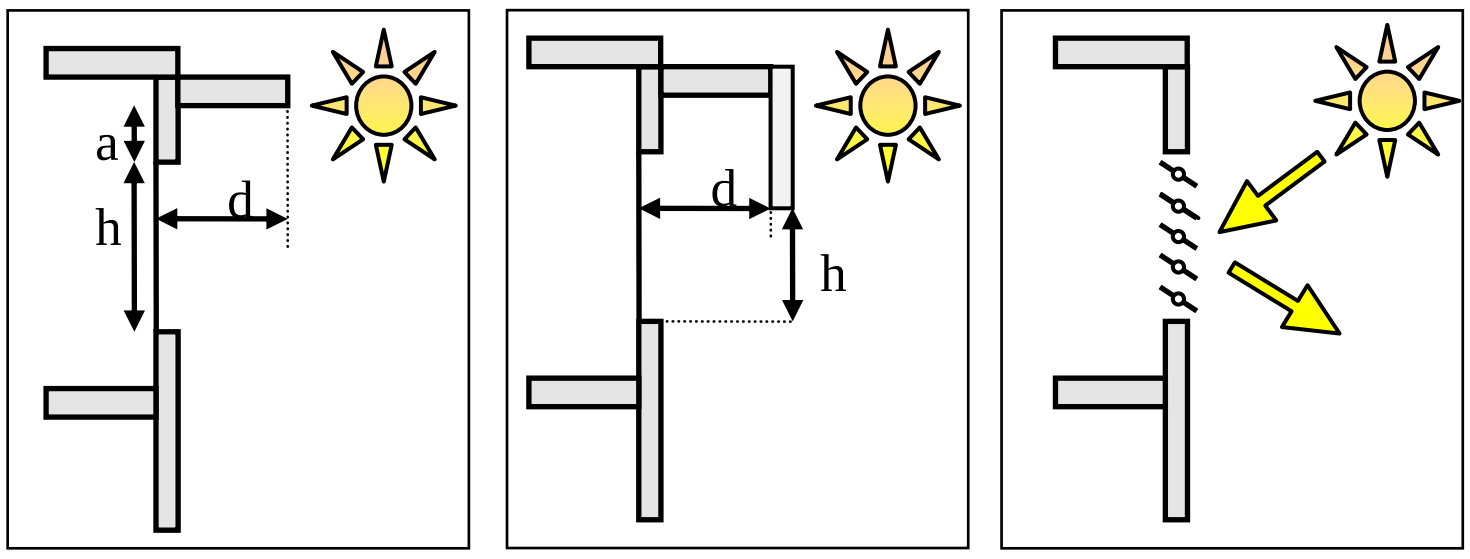

Fig. 3 : Some of the shading devices required by ECODOM

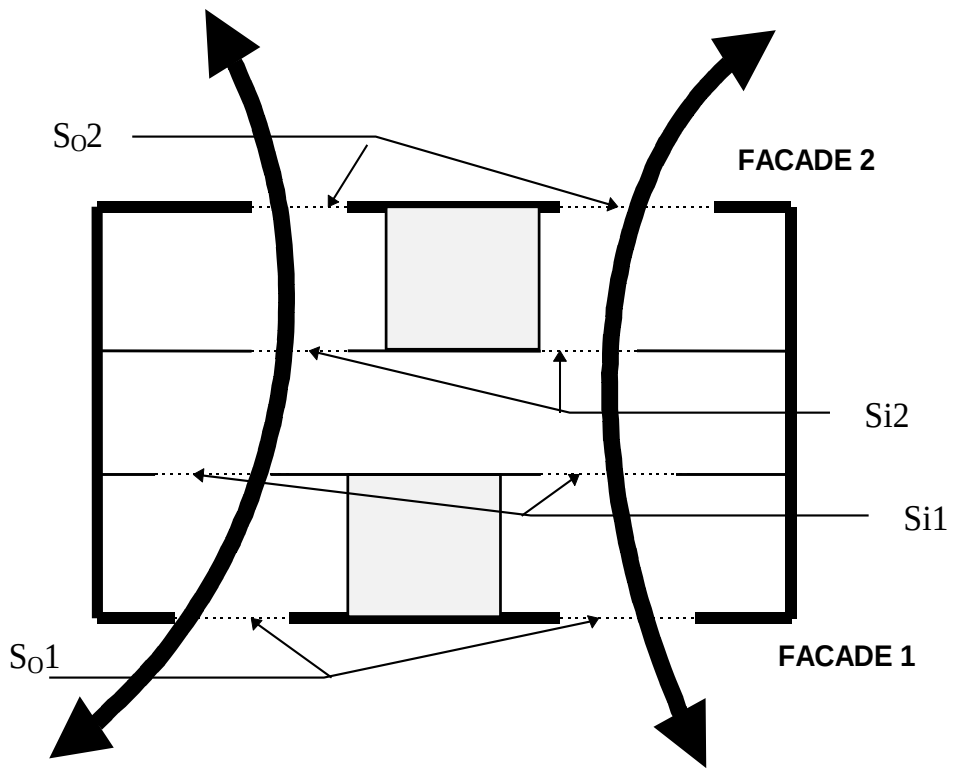

Fig. 4 : Cross ventilated dwelling

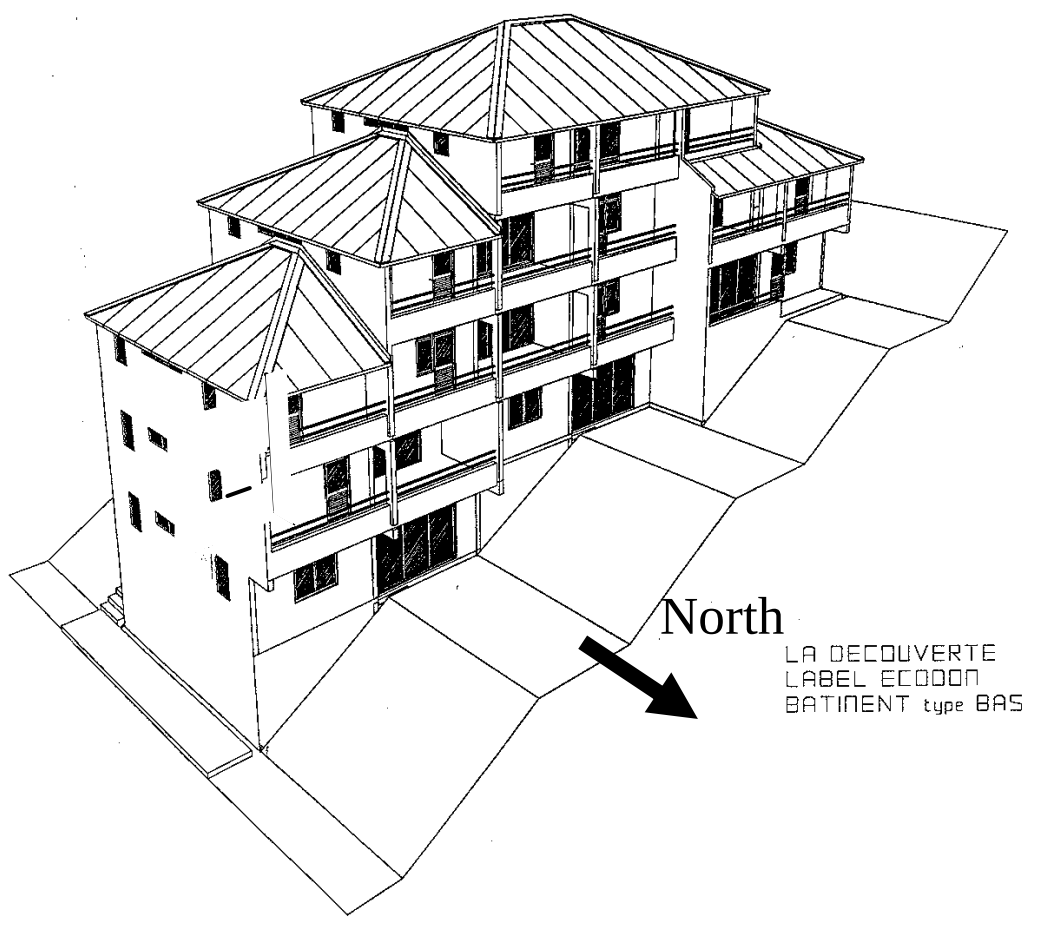

Fig. 5 : ECODOM “La Découverte”  - The original project (1996)

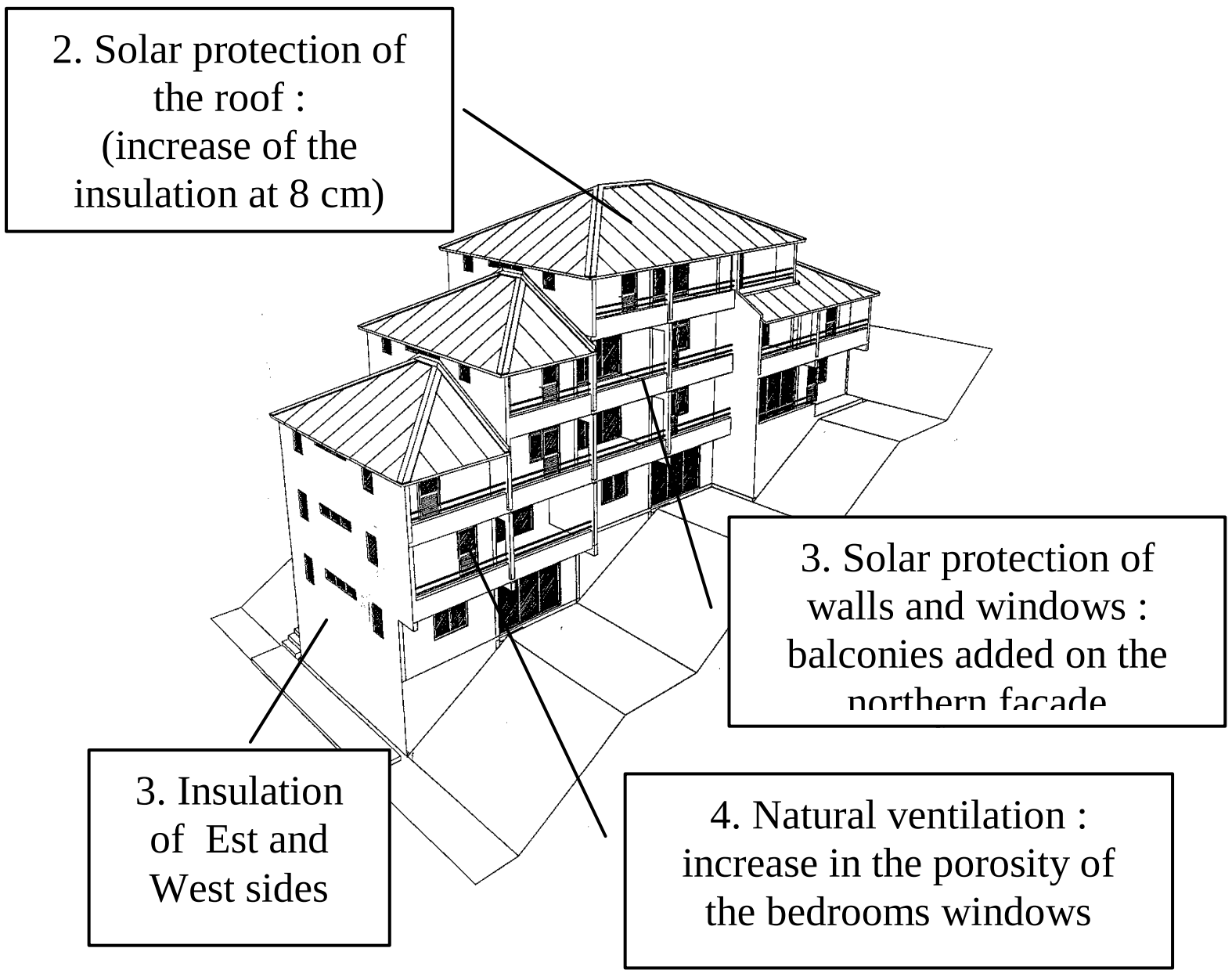

Fig. 6 : Modifications of the original project (1997)

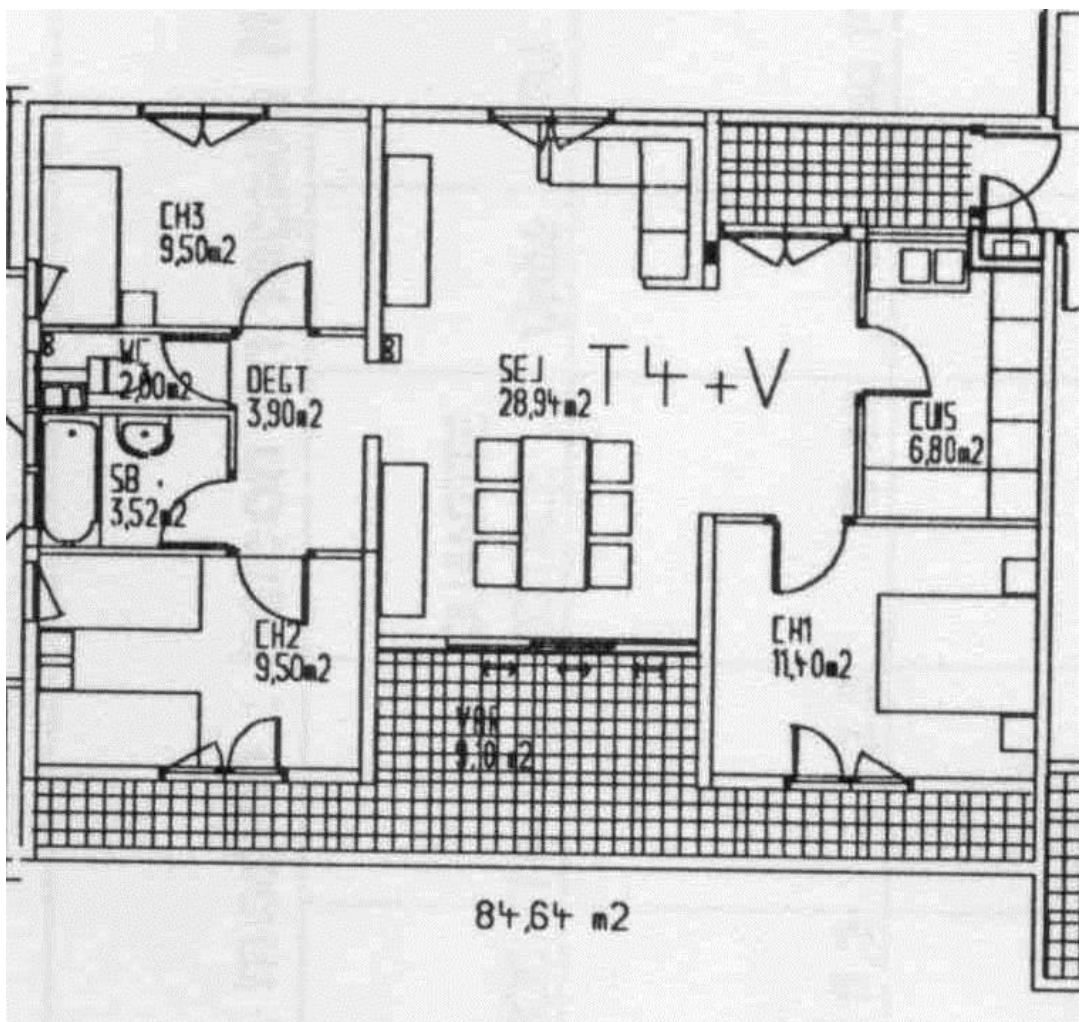

Fig. 7 : Floor plan of an ECODOM dwelling. Increase in porosity of bedrooms (glazed windows ) and solar protection of northern facade (balconies).

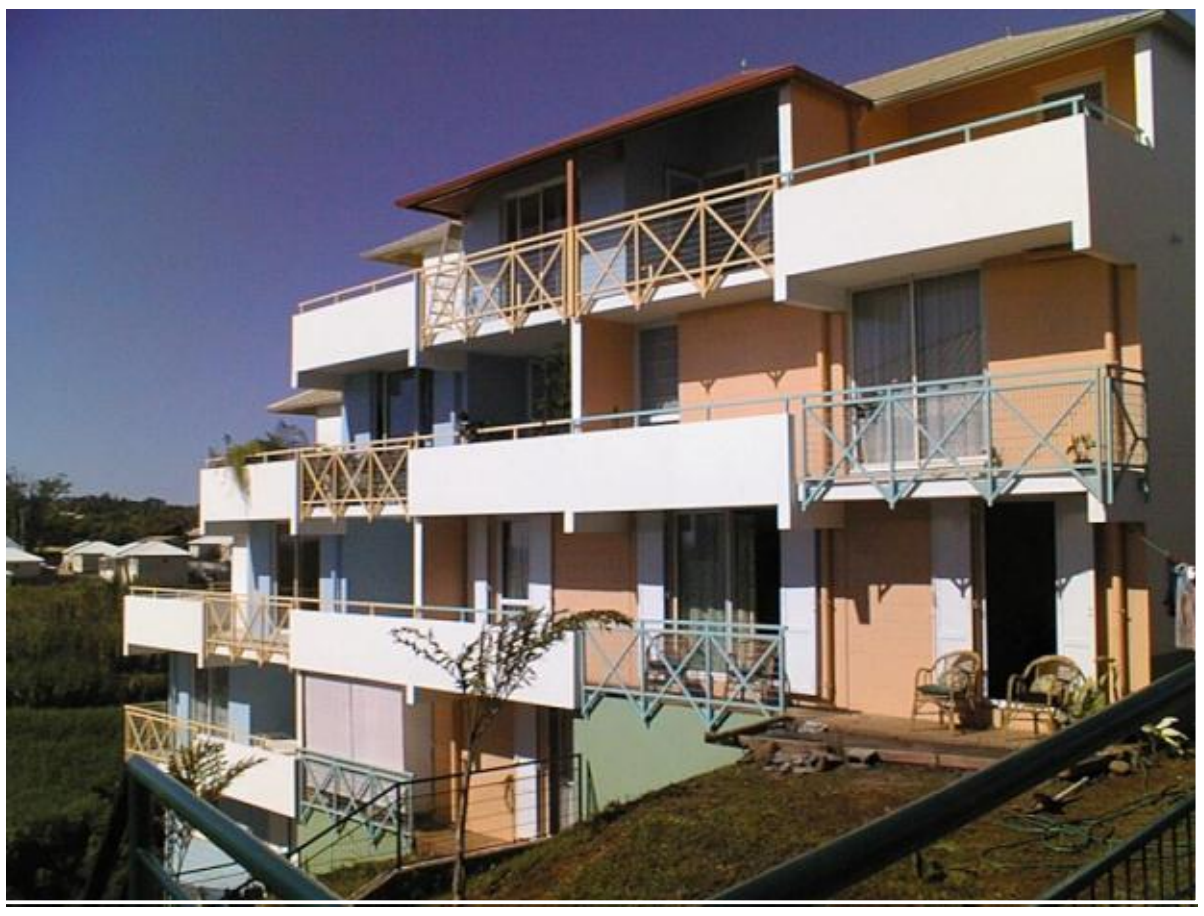

Fig. 8 : The final project (april 1999)

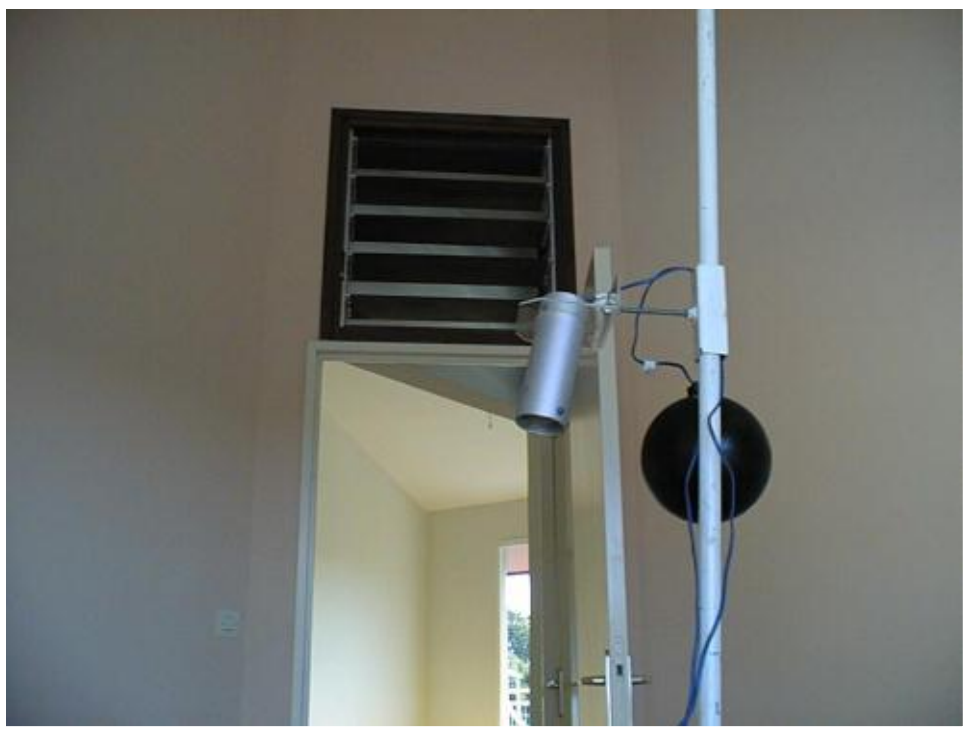

Fig. 9 : Measurements of ambient air and resultant temperature inside dwelling

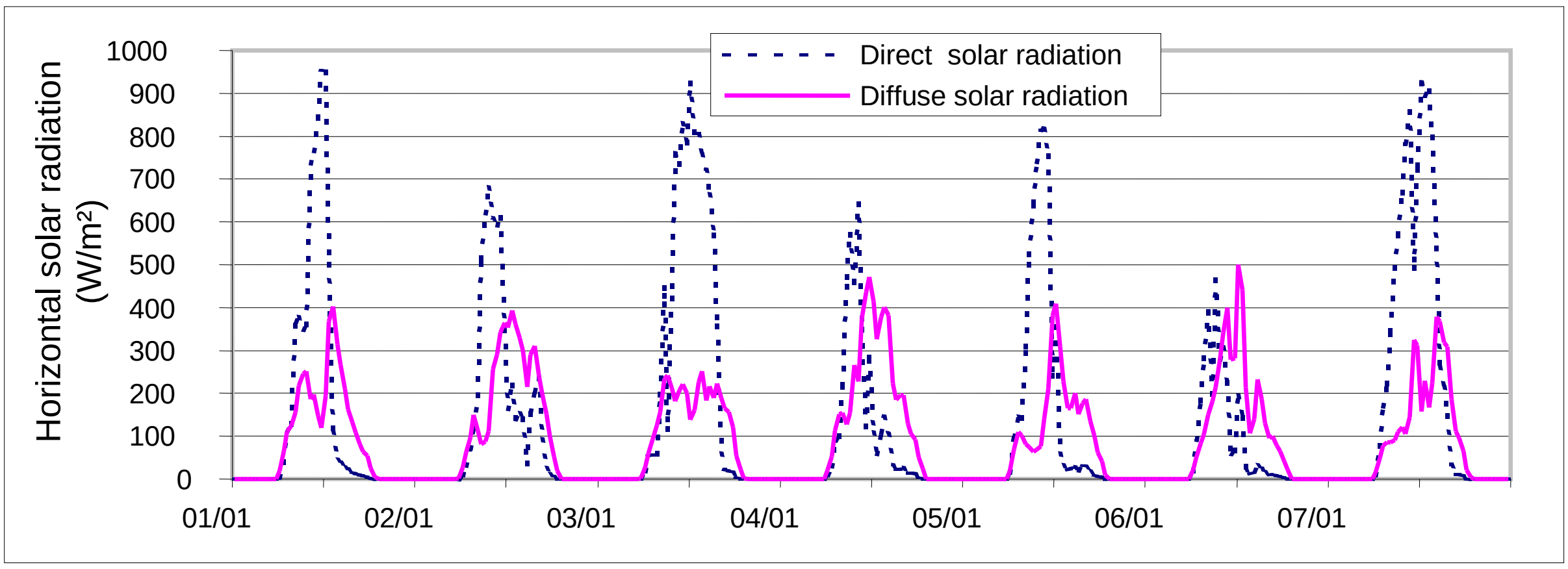

Fig. 10 : Solar radiation during the assessment of the solar protection of the roof

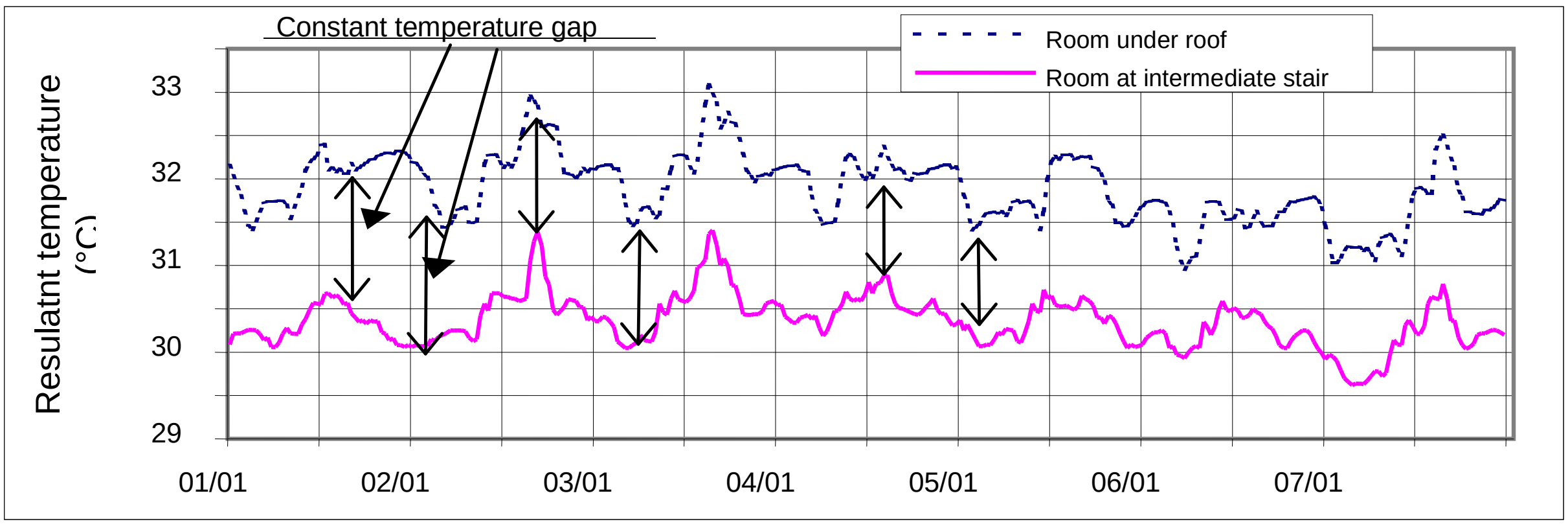

Fig. 11 : Comparison of the resultant temperature in two dwellings (one under roof and the other one at intermediate level

25

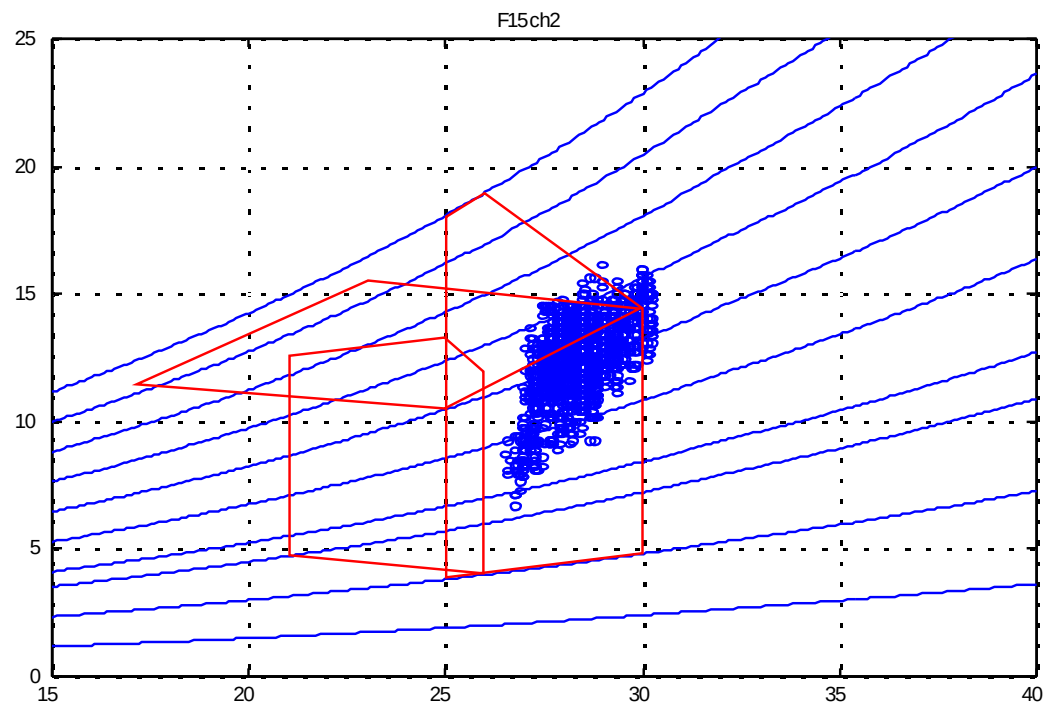

Fig. 12 : Temperature/humidity couples of an occupied flat, room at intermediate level

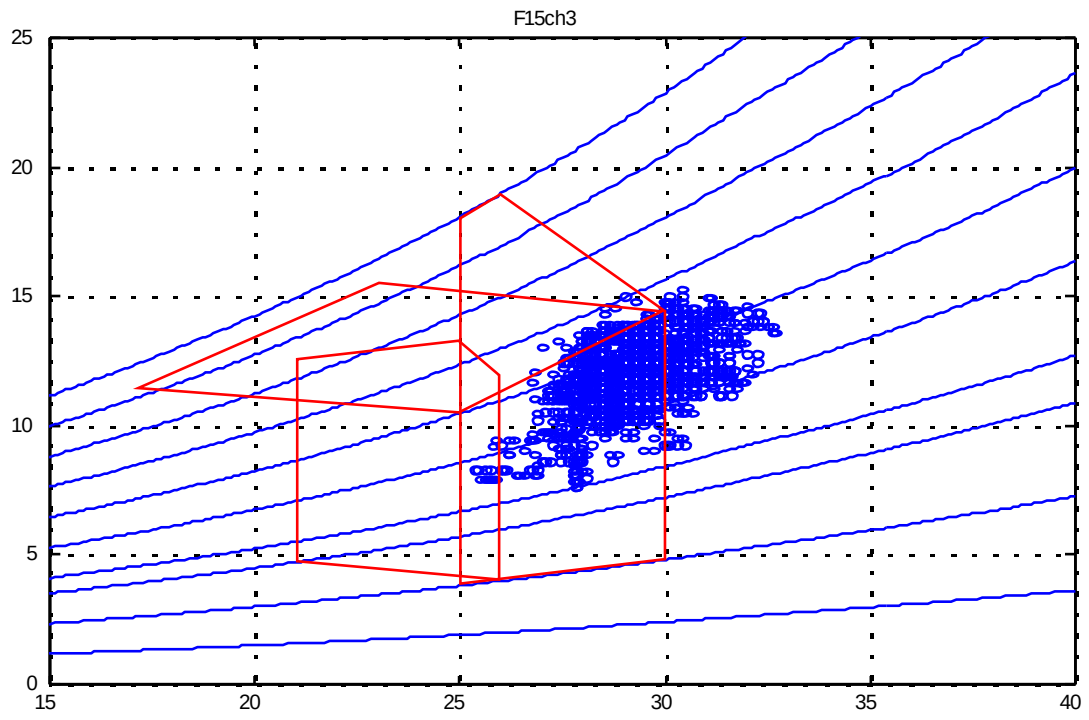

Fig. 13 : Temperature/humidity couples of an occupied flat, room under roof

Table 1 : Roof solar protection

| *Insulated simple roofs* | | |
|---|---|---|
| Roof colour | Polystyrene type insulation $\lambda$ = 0.041 W/m.K | Polyurethane type insulation $\lambda$ = 0.029 W/m.K |
| light ($\alpha$ = 0.4) | 5 cm | 4 cm |
| medium ($\alpha$=0.6) | 8 cm | 6 cm |
| dark ($\alpha$ = 0.8) | 10 cm | 8 cm |
| *Roofs with well-ventilated attics* | | |
| Roof colour | Polystyrene type insulation $\lambda$ = 0.041 W/m.K | Polyurethane type insulation $\lambda$ = 0.029 W/m.K |
| light ($\alpha$ = 0.4) | No insulation needed | |
| medium or dark ($\alpha$=0.6) | 2 cm | 0 cm |

Table 2 : Overhang - minimum d/h ratio values to be respected.

| | light colour | | | | medium colour | | | |
|---|---|---|---|---|---|---|---|---|
| Type of wall | East | South | West | North | East | South | West | North |
| Poured concrete 15cm (R=0.1 m².K/W) | 0.4 | 0.2 | 0.7 | 0.5 | 1 | 0.5 | 1.3 | 0.7 |
| hollow concrete blocks (R=0.2 m².K/W) | 0.1 | 0.1 | 0.3 | 0.2 | 0.5 | 0.3 | 0.8 | 0.5 |
| wood (R=0.5 m².K/W) | 0 | 0 | 0 | 0 | 0 | 0 | 0.2 | 0.1 |

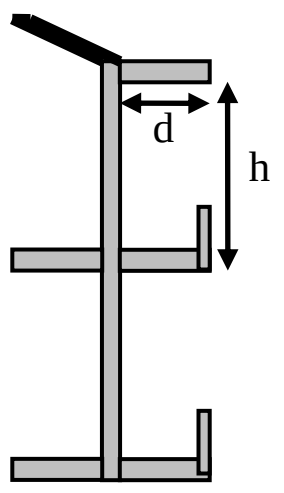

Table 3

Insulation of walls (in cm) for different orientations and external colours (for a conductivity of 0.041 W/m.K)

| | light colour | | | | medium colour | | | |
|---|---|---|---|---|---|---|---|---|
| Type of wall | East | South | west | North | East | South | West | North |
| concrete 20cm (R=0.1 m².K/W) | 1 | 1 | 1 | 1 | 2 | 1 | 2 | 2 |
| hollow concrete blocks (R=0.2 m².K/W) | 1 | 1 | 1 | 1 | 1 | 1 | 2 | 2 |
| wood (R=0.5 m².K/W) | 0 | 0 | 0 | 0 | 0 | 0 | 1 | 1 |

Table 4

Values of d/(2a+h) (case 1), or d/h (case 2)

| | Orientation of windows | | | |
|---|---|---|---|---|
| | East | South | West | North |
| Reunion Island | 0.8 | 0.3 | 1 | 0.6 |

Table 5 : Technical characteristics - solar water heaters

| Solar water heaters minimum collector area to be installed for various types of dwellings | |
|---|---|
| F1-F2 (2 rooms) | 1.5 m² |
| F3 (3 rooms) | 2.0 m² |
| F4 (4 rooms) | 2.5 m² |
| F5 (5 rooms) | 3.0 m² |
| F6 and more | 3.5 m² |

Table 6 : Experimental schedule for non occupied and occupied periods

| | Non-occupied period | Occupied period |
|---|---|---|
| « La Découverte » | From january to april, 1999 | No measurement |
| « La Trinité » | From january to may,1998 | From march to april, 1998 |